\documentclass[prl,twocolumn,superscriptaddress]{revtex4-1}
\usepackage{epsfig}
\usepackage{graphicx}
\usepackage{palatino}
\usepackage[english]{babel}
\usepackage{hyphenat}
\usepackage{amsmath}
\usepackage{amssymb}
\usepackage{slashed}
\usepackage{epstopdf}
\usepackage{xcolor}
\usepackage{booktabs}
\definecolor{lcolor}{rgb}{0.,0.0,0.}
\definecolor{citcolor}{rgb}{0,0.,0.5}
\usepackage[breaklinks,colorlinks,urlcolor=blue,citecolor=blue,linkcolor=blue]{hyperref}
\usepackage{multirow}
\usepackage{ltablex}

\def\bs{\boldsymbol} 
\def\del{\partial}
\def\bdel{\bs\partial}

\newcommand{\eqn}[1]{Eq.~\eqref{#1}}
\newcommand{\fign}[1]{Fig.~\ref{#1}}
\long\def\comment#1{ }

\newcommand{\nn}{\nonumber\\ }

\def\be{\begin{eqnarray*}}
\def\ee{\end{eqnarray*}}
\def\beq{\begin{eqnarray}}
\def\eeq{\end{eqnarray}}

\def\bwt{\begin{widetext}}
\def\ewt{\end{widetext}}
\newcommand{\bea}{\beq \begin{aligned}}
\newcommand{\eea}{\end{aligned}\eeq}

\def\k{{\boldsymbol k}}

\def\bell{{\boldsymbol \ell}}
\def\v{{\boldsymbol v}}
\def\z{{\boldsymbol z}}

\def\0{{\boldsymbol 0}}

\def\k{{\boldsymbol k}}
\def\x{{\boldsymbol x}}
\def\y{{\boldsymbol y}}

\def\r{{\boldsymbol r}}
\def\b{{\boldsymbol b}}
\def\A{{\boldsymbol A}}

\def\rme{{\rm e}}
\def\rmd{{\rm d}}

\def\scal{\text{scal}}

\def\tr{ \text{Tr}}

\def\cG{{\cal G}}
\def\cU{{\cal U}}
\def\cO{{\cal O}}
\def\cP{{\cal P}}

\begin{document}
\title{{\bf  A novel formulation of the unintegrated gluon distribution for DIS }}
\author{Renaud Boussarie}
\email[]{rboussarie@bnl.gov}
\affiliation{Physics Department, Brookhaven National Laboratory, Upton, NY 11973, USA.}
\author{Yacine Mehtar-Tani}
\email[]{mehtartani@bnl.gov}
\affiliation{Physics Department, Brookhaven National Laboratory, Upton, NY 11973, USA.}
\affiliation{RIKEN BNL Research Center, Brookhaven National Laboratory, Upton, NY 11973, USA}

\begin{abstract} 
We provide a semi-classical description of the inclusive gluon induced Deep Inelastic Scattering cross section in a way that accounts for the leading powers in both the Regge and Bjorken limits. Our approach thus allows a systematic matching of small and moderate $x_{\rm Bj}$ regimes of gluon proton structure functions. We find a new unintegrated gluon distribution with an explicit dependence on the longitudinal momentum fraction $x$ which entirely spans both the dipole operator and the gluonic Parton Distribution Function. Computing this gauge invariant gluon operator on the lattice could allow to probe the energy dependence of the saturation scale from first principles.
 \end{abstract}
\keywords{Perturbative QCD, DIS, Saturation, small-x, gluon}
\maketitle

\section{Introduction}

At asymptotically short distances the proton behaves as a collection of free quarks and gluons (partons). This regime of QCD is probed for example in Deep Inelastic Scattering (DIS) experiments such as electron-proton collisions, where a highly virtual photon of momentum $q$ and virtuality $Q^2=-q^2 \gg \Lambda_{\rm QCD}^2$ is exchanged between the electron and the hadronic target. At small  $x_{\rm Bj }\equiv Q^2/2(q\cdot P)$, where $P$ is the 4-momentum of the proton, the number of gluons probed in the proton rises rapidly and it is expected to reach saturation at very high energies due to gluon recombination effects. This takes place at the saturation scale $Q_s$~\cite{glr,mq}, which increases with decreasing $x_{\rm Bj}$. This remarkable yet elusive emergent phenomenon is the subject of active experimental research.

The probability for a parton to carry a fraction $x$ of the proton momentum, known as the parton distribution function (PDF), is encompassed by the structure functions probed in DIS. PDF's obey renormalization group equations, the Dokshitzer-Gribov-Lipatov-Altarelli-Parisi (DGLAP) equations~\cite{dglap}, which appear in the Bjorken limit, $Q \rightarrow \infty$ at fixed $x_{\rm Bj}$. In that limit, an expansion in powers of $1/Q$ separates short-distance physics, i.e. the hard subprocess, from the long-distance physics encoding the non-perturbative dynamics of confinement in the proton via the PDF. To leading logarithmic accuracy the dominant contribution arises from diagrams that connect the target to the photon with a strong ordering from small to large transverse momenta $\mu \ll k_{\perp,1} \ll k_{\perp,2} \ll ... \ll Q$, while the longitudinal components along the dominant light cone direction of the proton are of similar magnitude: $x_{\rm Bj}P^+ \sim k_1^+ \sim  k_2^+ \sim ...\sim P^+$, where $P^+$ is the large momentum component of the proton  \footnote{It is customary in collider physics to use light cone variables defined as $k^+=(k_0+k_3)/\sqrt{2}$ and $k^-=(k_0-k_3)/\sqrt{2}$ in addition to the transverse components $\k \equiv (k^1,k^2)$. Hence, in the frame where the proton is moving in the $+z$ direction  its dominant momentum component is $P^+$.}. 

In addition to the Bjorken limit, attention has been given to the Regge limit, $x_{\rm Bj}\rightarrow 0$ at fixed $Q$. Here, an expansion in powers of $x_{\rm Bj}$ is to be performed and gluon saturation is expected to emerge for very small values of this variable.

In DIS, the dominant process at small $x_{\rm Bj}$ is that of the virtual photon splitting into a quark-antiquark dipole which subsequently interacts with the target by the $t$-channel exchange of gluons with small longitudinal momenta in the light cone direction of the photon. In this case, the quantum mechanical time associated with the dipole formation, $(x_{\rm Bj} P^+)^{-1}$ is much larger than the target longitudinal size of order $1/P^+$. This opens up a large phase space for long lived quantum fluctuations that are enhanced by potentially large logarithms of the form $\alpha_s \log 1/x_{\rm Bj} \sim 1$. The latter  are generated by strong ordering in the exchanged longitudinal gluon momenta $k_i^-$. In contrast to DGLAP evolution, the transverse components are assumed to be of the same order, i.e., $k_{\perp 1} \sim k_{\perp 2} \sim ...\sim k_{\perp n}$ along the cascade. This implies a strong inverse ordering the $k^+_i\sim k^2_{i \perp}/k^-_i$ components. 

However, it was observed in Ref.~\cite{negativeXS} that the Next-to-Leading Order (NLO) corrections (see~\cite{nlobfkl,nlobk,nlojimwlk}) to the Balitsky-Fadin-Kuraev-Lipatov (BFKL)~\cite{bfkl}, Balitsky-Kovchegov (BK)~\cite{bkian,bkyuri} and Jalilian-Marian-Iancu-McLerran-Weigert-Leonidov-Kovner (JIMWLK)~\cite{jimwlk} equations that govern small-$x_{\rm Bj}$ physics, generate large collinear logarithms that have to be resummed to insure numerical stability of the equations~\cite{collinear-logs-Beuf,collinear-logs-Edmond}.

The origin of this problem can be traced back to the so-called shock wave approximation which assumes that the target longitudinal extent, as perceived by the photon, is equal to zero. For instance two successive gluons with $k^-_1\gg k_2^-$ should also obey the following ordering $k^+_1\ll k_2^+$. However, the transverse integrations are not explicitly constrained through the evolution,  resulting in the extension of the $k^+$ phase-space in the non-physical region $k^+_1 > k_2^+$. More importantly, when $k^+_1 \sim k_2^+$, with $k^-_1\gg  k_2^-$ we have $k_{\perp 1} \gg  k_{\perp 2}$, which corresponds to the DGLAP region. 

In light of these recent developments, one can reasonably hope that a reformulation of small $x$ physics may cure this problem without resorting to an order by order resummation of secular terms. This is the goal of the present letter. 

We will use a semi-classical approach~\cite{mv} similar to the shock wave approach~\cite{bkian} but we will refrain from making assumptions about the extent of the target. The leading powers in both the Bjorken and the Regge limits are instead obtained by performing a gradient expansion around the transverse position of quantum fluctuations.

This Letter is organized as follows.  We first present the operator definition for a novel unintegrated gluon distribution that appears in inclusive DIS within our approach, and which interpolates between the Bjorken and Regge limits. Then we explicitly derive the general DIS cross section in our scheme. We finally perform the classical expansion to recover the interpolating expression for both kinematic limits and show how the aforementioned gluon distribution appears in inclusive DIS.

\section{Unintegrated gluon distribution function at small x and beyond: operator definitions}
Correcting for the $x$ dependence of the dipole scattering amplitude yields a novel unintegrated gluon distribution which shall be derived in the context of DIS in the next section. It reads
\bwt
  \beq\label{eq:dist-def}
&&xG^{ij}(x,\k)  = \int \frac{\rmd \xi^- \rmd \r }{(2\pi)^3 P^+}\, \rme^{i xP^+\xi^-  -i \k\cdot \r} \int_0^1 \rmd s \int_0^1 \rmd s'
 \,\langle P|  \, \tr  \,  \cU_\0(s\r,s'\r) F^{i+}(\xi^-,s'\r) \, \cU_\r(s'\r,s\r) F^{j+}(0,s\r) | P \rangle \,,\nn
 \eeq
 \ewt
where  $\k$ is a transverse momentum and $i,j=1,2$ label two orthogonal transverse directions. Also, $F^{i+}=\del^iA^+-\del^+A^i-ig[A^i,A^+]$ is the field strength tensor
and 
 \beq
 \cU_\0(s\r,s'\r) &=&  [s\r,\0]_{0^-}[0,\xi^-]_\0[\0,s'\r]_{\xi^-}\,, \nn
 \cU_\r(s'\r,s\r) &=&  [s'\r,\r]_{\xi^-}  [\xi^-,0]_\r[\r,s\r]_{0^-}\,,  
 \eeq
 are two finite length staple-shaped gauge links that connect  $F^{i+}(\xi^-,s'\r)$ to $F^{i+}(0^-,s\r)$ as depicted in Fig.~ \ref{fig:splitting}~\footnote{$F^{i+}(\xi^-,\r)\equiv F^{i+}(\xi^-,\xi^+=0,\r)$}, and where
\beq
 [\xi^-,0^-]_{\r}&& \equiv  \cP \exp\left[ig \int^{\xi^-}_{ 0^-} \rmd x^-A^+(x^-,\r)\right]
\eeq
and
\beq
 [ \x,\y]_{\xi^-} \equiv \cP \exp\left[-ig \int_{ \y}^{ \x} \rmd \z \cdot  \A(\xi^-,\z)\right]
\eeq
are path ordered Wilson lines in the $+$ and $\perp$ directions, respectively, with $\z \equiv \z(s) = s \,\x +(1-s)\,\y$. 

One can readily verify that the distribution (\ref{eq:dist-def}) encompasses both the gluon PDF at large $x$ and the dipole unintegrated distribution at small $x$. 
\begin{figure}
\begin{center}
\includegraphics[width=0.4\textwidth]{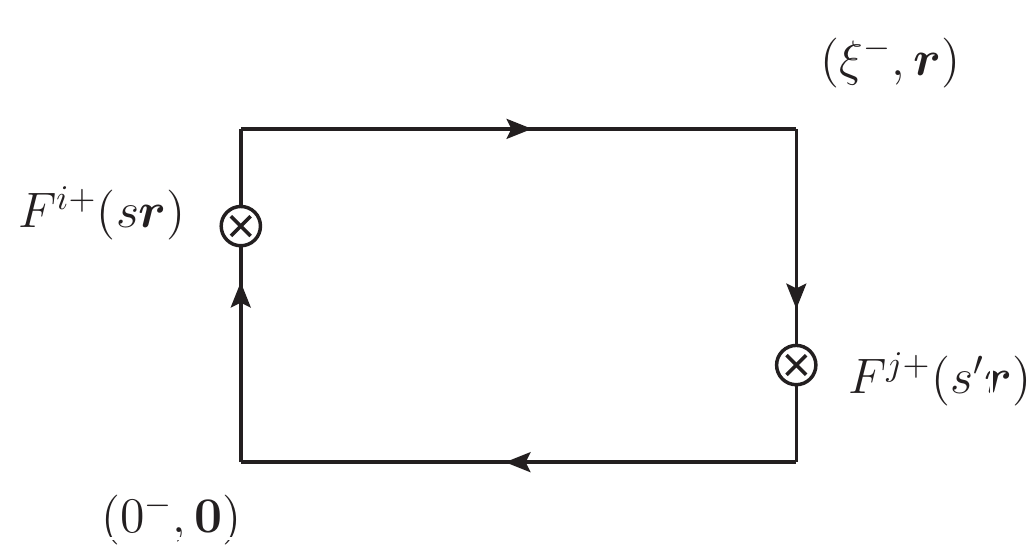} 
\end{center}
\caption{Diagrammatic depiction of the nonlocal operator that defines the unintegrated gluon distribution  in \eqn{eq:dist-def}. The horizontal and vertical lines represent path ordered Wilson lines along the $+$ and transverse directions, respectively.  }
\label{fig:splitting}
\end{figure}
Integrating over $\k$ yields a $\delta(\r)$ and one recovers the gluon PDF
  \beq\label{eq:dist-def-pdf}
&& \int \rmd^2 \k \, x G^{ii}(x,\k)  =xg(x) \equiv \int \! \frac{\rmd \xi^-}{ (2\pi)P^-} \rme^{i xP^+\xi^- }  \, \nn
 &&\times \,\langle P|  \, \tr  \, [0,\xi^-] F^{i+}(\xi^-) [\xi^-,0] F^{i+}(0) | P \rangle\,,
 \eeq
where the gluonic operator is implicitly evaluated at $\r=0$. The dipole scattering amplitude relevant at small $x$ is obtained from  \eqn{eq:dist-def}  by first setting $x=0$, and neglecting the transverse gauge links that can be gauged away along with all $A^i$ fields. One can get a more symmetric form for the operator by using translational invariance and the fact that $\int \rmd \b\, \rmd \zeta^-= (2P^+)^{-1}\langle P |P\rangle$) to write $0 \rightarrow \zeta^-$ and $\0 \rightarrow \b$ with the price of the proton normalization in the denominator. Then one can notice that $\int_0^1 \rmd s F^{i+}(s\r)=\int_0^1 \rmd s \,\del^i A^+(s\r) =  \frac{r^i}{\r^2} [ A^+(\r)-A^+(\0) ]$ and similarly for $\int_0^1 \rmd s \, r^i F^{i+}(s'\r)$. These differences of $A^+$ terms result from taking the derivative of the operator  $\tr [\xi^-, \zeta^-]_\r [\zeta^- , \xi^-]_\0 $ w.r.t. $\xi^-$ and $\zeta^-$, respectively. Upon integration over $\xi^-$ and $\zeta^-$, one finally obtains
 \begin{equation}
  x G^{ij}(x,\k) \rightarrow \int \!  \frac{\rmd \b \,\, \rmd \r}{(2\pi)^4} \frac{ \rme^{-i \k \cdot \r}}{\alpha_s} \frac{r^i r^j }{\r^4}\, 
 \frac{\langle P|\,  \tr \, U_{\b+\r} U^\dag_{\b} -N_c |P\rangle }{\langle P |P\rangle }\,
 \end{equation} 
 where $U_\r = [+\infty,-\infty]_\r$. This result is compatible with the definition of the unintegrated gluon distribution at small $x$, which yields a form of the BK equation that is local in momentum space, see e.g.~\cite{Kovchegov:1999ua,Munier:2003vc,updf}

\section{DIS beyond the shock wave approximation}
In this section we will demonstrate how the unintegrated gluon distribution introduced in the previous section emerges in a physical observable, namely, inclusive DIS, but first let us summarize the three main approximations that will be made in the following derivation. First of all, we will only focus on gluon contributions to the cross section, since we want to improve a small-$x$ inspired scheme where gluons dominate. Secondly, we adopt $k^-$ as a factorization variable between the target fields and the quantum fluctuations which allows to deal with powers of $s\to \infty$ without any further specification on $Q^2$. It allows to resum both collinear and rapidity  logarithms when $Q^2 \sim s$  and $Q^2 \ll s$, respectively~\cite{Balitsky:2015qba}. We will initially restrict ourselves to these two assumptions, but eventually a classical expansion in powers of an intrinsic transverse momentum in the proton, $k_\perp /\sqrt{s}$, will be performed.

Consider the DIS subprocess $\gamma^\ast(q) +{ \rm proton}\,(P)  \to  X$. Owing to the optical theorem, the total cross-section is related to the forward scattering amplitude $\gamma^\ast(q)+P \to \gamma^\ast(q)+P$. We shall use  light cone variables  $(k^+,k^-,\k)$, defined by the 4-vector decomposition
$k^\mu = k^+  n^\mu  + k^- \bar n^\mu + k_\perp^\mu\,$, 
where the light cone vectors $n$ and $\bar n$ satisfy $n^2=\bar n^2=0$ and $n\cdot \bar n=1$. We choose the frame in which the photon and proton momenta are aligned with the $z$ axis.
The leading contributions in both Regge and Bjorken limits stems from gluons with a negligible $-$ component of their momentum. This reflects itself in coordinate space as an expansion on the null plane $x^+=0$, which implies for the target gluon field $A^{\mu} \simeq A^{\mu}(0,x^-,\x)$ only depends on $x^-$ and the transverse coordinate $\x$. This leading field is generated by a color charge current whose only non-vanishing component is $J^+(x^-,\x)$. It is straightforward to see that the Yang-Mills equations in covariant gauge admit the solution $A^+(x^-,\x)$ and $A^-=A^i=0$ where $-\bdel^2 A^+(x^-,\x) = J^+(x^-,\x)$ \cite{GMT}. This solution is also common to the light cone gauge $A^-=0$.  A more general solution can be obtained by an arbitrary gauge rotation $\Omega(x^-,\x)$ which generates a transverse pure gauge field $ig A^i =- \Omega\del^i \Omega^{-1} $. In the shock wave approximation, one would assume that the current is very peaked around $x^-=0$ and build effective Feynman rules by expanding around this point~\cite{next-to-eikonal,Chirilli:2018kkw}. Here, we will refrain from using this assumption from the get-go, and we will instead perform a gradient expansion in transverse position space in the final expressions for the cross-section.

\begin{figure}
\begin{center}
\includegraphics[width=0.4\textwidth]{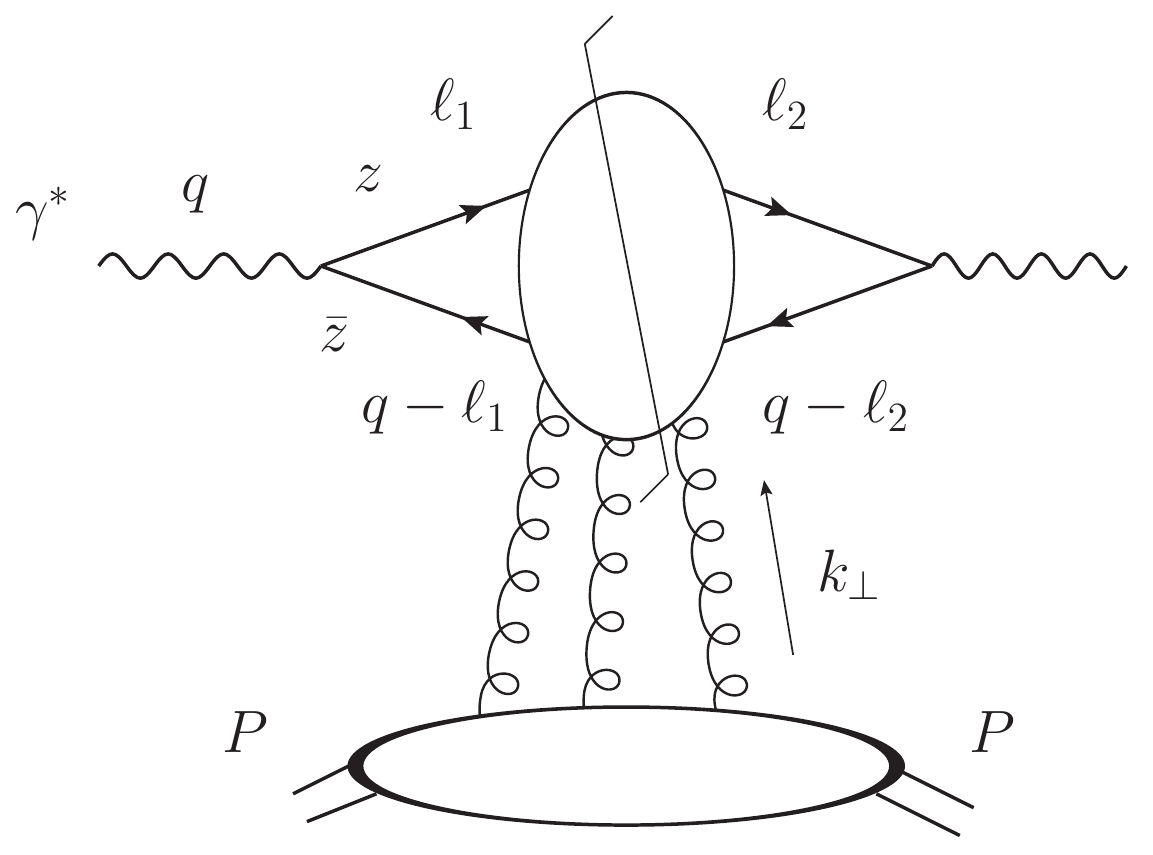} 
\end{center}
\caption{ Diagrammatic representation of the process $\gamma(q) + A  \to q  \, \bar q$. }
\label{fig:splitting}
\end{figure}

The transverse and longitudinal cross-sections are related to the hadronic tensor as follows
$\sigma_T(x,Q^2) =\frac{2 \pi m}{s-m^2}  e^2 \varepsilon^{\ast \mu}_{\lambda}  W_{\mu\nu} \varepsilon^{\nu}_{\lambda}$, with $\lambda=\pm 1 $ and $\sigma_L(x,Q^2) =\frac{2 \pi m}{s-m^2}  e^2 \varepsilon_L^{\ast \mu} W_{\mu\nu} \varepsilon^{\nu}_L$ \,.
The longitudinal polarization vector may be chosen to be $\varepsilon^{\mu}_L =\frac{1}{Q} \left( q^- n +\frac{Q^2}{2q^-}\bar n\right)\,$,
while the transverse polarizations satisfy $\varepsilon_{+1} \cdot \varepsilon^\ast_{-1} =0 \, \quad \text{ and}  \quad \sum_{\lambda=\pm 1} \varepsilon^\mu_\lambda \cdot \varepsilon^{\ast \nu}_\lambda = g_\perp^{\mu\nu}$.

The first and second working assumptions, namely considering a gluonic target boosted on the light cone, allow us to write the hadronic tensor in the form  $\tr \left[ \gamma^\mu D_F(\ell_2,\ell_1) \gamma^\nu  D_F( q-\ell_1,q-\ell_2)\right]$, where $D_F$ is the Dirac propagator in the target background field $A^+$. See \fign{fig:splitting}.

Because the background (target) field does not depend on $x^+$,  its Fourier transform is proportional to $\delta(k^-)$. As a result, the - components of the quark and antiquark momenta are conserved. This implies the following Dirac decomposition for the quark propagator in momentum space:
\begin{equation}\label{eq:dirac-p-mom}
D_F(\ell_2,\ell_1)= \frac{i\gamma^-}{2\ell_1^-} (2\pi)^4\delta(\ell_2-\ell_1)+ \frac{\slashed{\ell}_2 \gamma^- \slashed{\ell}_1}{2\ell_1^-} G_\scal(\ell_2,
\ell_1)\,,
\end{equation}
where the scalar propagator obeys the Klein-Gordon equation $\left( -\Box_x+ 2 ig  A^+(x)\del^-_x\right) G_\scal(x,x_0)= \delta(x-x_0)$. 
The instantaneous term of the propagator, i.e. the first term in the r.h.s. of \eqn{eq:dirac-p-mom}, does not contribute to the DIS cross section. We are therefore left with the scalar propagator term. 

Making use of its independence on $x^-$, the scalar propagator can be expressed as follows in Schwinger notations:
\beq
&& G_\scal(\ell_2,\ell_1)=\frac{2\pi }{2i\ell_1^-}\,\delta(\ell_2^--\ell_1^-)\, \int \rmd \xi_1^-\int  \rmd \xi_2^-\,\nn &&\,\times (\bell_2| \cG_{\ell_1^-}(\xi_2^-,\xi_1^-)|\bell_1)\, \rme^{i\ell_2^+ \xi_2^--i \ell_1^+ \xi_1^- }\,.
\eeq
$\cG_{\ell_1^-}(\xi_2^-,\xi_1^-)$ is nothing but the propagator of a non-relativistic particle in 2+1 dimension. It satisfies the Schr\"{o}dinger equation for the final time
\begin{equation}
\left[ i\frac{\partial}{\partial\xi_2^-} - \frac{\hat{\boldsymbol{P}}^2}{2\ell^-} +g A(\xi_2^-,\hat{\x}) \right] \cG_{\ell^-}(\xi_2^-,\xi_1^-) = i\delta(\xi_2^--\xi_1^-)\,,
\end{equation}
and a similar equation for the initial time.
Here, $\hat{\boldsymbol{P}} =  i \boldsymbol{\partial}$ is the momentum operator. The free propagator is recovered when setting $A^+ = 0$ in these equations, and reads  $\cG^{(0)}_{\ell^-}(\xi^-_2,\xi_1^-)=\rme^{-i \hat{\boldsymbol{P}}^2 /2\ell^- (\xi_2-\xi_1)^-}$. For more insight about the $\cG$ operator, the reader is refered to~\cite{Blaizot:2015lma}.

With the help of the effective propagators, non-trivial algebra and multiple uses of the integral form of the Schr\"{o}dinger equation, the cross section can be cast into:
\beq\label{eq:cross-section-position}
 && \sigma =8\alpha_{s}\alpha_{\mathrm{em}}\sum_{f}q_{f}^{2}\,\mathrm{Re}\int_{0}^{1}\!\frac{\rmd z}{2\pi}\int\!\rmd x_2^-\rmd x_1^-\rmd^{2}\r\,\rmd^{2}\r^{\prime}  \nn
 && \times [ \, \varphi_{L}(\r)\,\varphi_{L}^{\ast}(\r^{\prime})+\frac{1}{2}\sum_{\lambda,\lambda^{\prime}}\,\varphi_{T}^{\lambda,h}(\r)\varphi_{T}^{\lambda^{\prime},h\ast}(\r^{\prime})\,]  \label{eq:XS} \\
 && \times \rme^{iq^{+}(x_{2}-x_{1})^{-}}\!\int\!\rmd^{2}\x_2\,\rmd^{2}\x_1\frac{\langle P|\mathcal{O}(x_{2}^{-},x_{1}^{-};\x_2,\x_1,\r,\r^{\prime})|P\rangle}{\left\langle P|P\right\rangle }.\nonumber
\eeq
This expression involves the longitudinal and transverse photon wave functions, respectively~(see e.g.~\cite{Beuf:2011xd}):
\begin{equation}
\varphi_{L}(\r)=2z\bar{z}QK_0(\sqrt{z\bar{z}Q^2\r^2}), \label{eq:phiL}
\end{equation}
and
\begin{equation}
\varphi_{T}^{\lambda,h}(\r)=iQ\sqrt{z\bar{z}}(z-\bar{z}+2\lambda h)\frac{(\boldsymbol{\epsilon}_{T}^{\lambda}\cdot\r)}{\left|\r\right|}K_1(\sqrt{z\bar{z}Q^2\r^2}), \label{eq:phiT}
\end{equation}
where $z=\ell_1^-/q^-$ is the loop quark's longitudinal momentum fraction, $\bar z =1-z$, and $h=\pm 1/2$ is the quark helicity. In the present scheme, the complicated dependence of the $t$-channel operator has the consequence that the wave functions are evaluated a different dipole sizes, contrary to what the dipole or shock wave framework would have led to. The operator in \eqn{eq:XS} reads:
\bwt
\beq
\mathcal{O}(x_2^-,x_1^-;\x_2,\x_1,\boldsymbol{r},\r^{\prime}) & = & \mathrm{tr}\left\{(\x_2|\mathcal{G}_{zq^{-}}(x_2^-,x_1^-)|\x_1)[A^+(x_1^-,\x_1+\r) -A^+(x_1^-,\x_1)] \right.\\
 & \times & \left. (\x_1+\r |\mathcal{G}_{-\bar{z}q^{-}}(x_1^-,x_2^-)|\x_2+\r^\prime) [ A^+(x_2^-,\x_2+\r^\prime)-A^+(x_2^-,\x_2) ]\right\}. \label{eq:op}
\eeq
\ewt

\begin{figure}
\begin{center}
\includegraphics[width=0.27\textwidth]{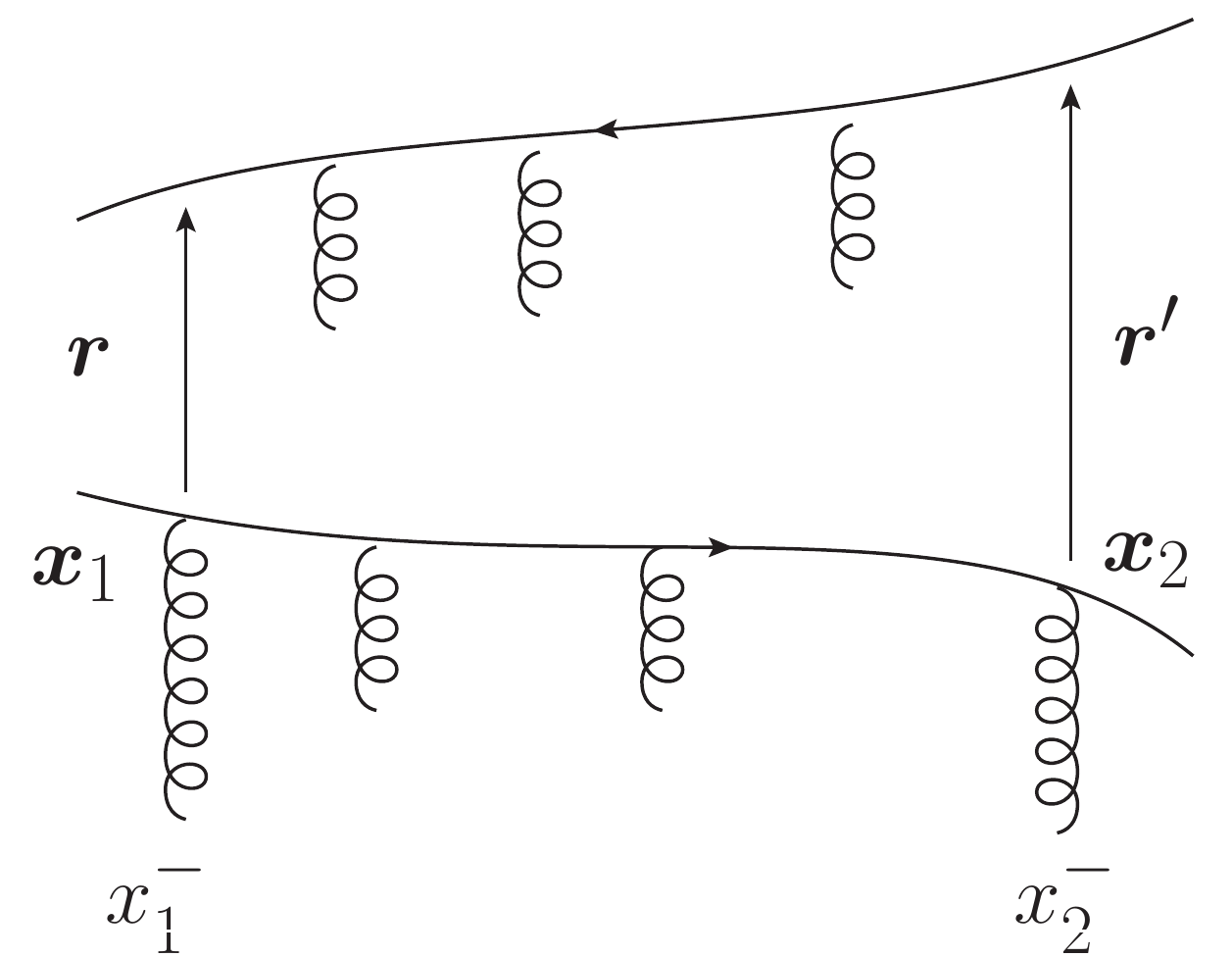} 
\end{center} 
\caption{One contribution to the operator from \eqn{eq:op}. The quark and the antiquark undergo the propagation in the external field between their first and final interaction, respectively at times $x_1^-$ and $x_2-$, here both on the quark at respective transverse positions $\x_1$ and $\x_2$.}
\label{fig:operator}
\end{figure}

It describes the Brownian motion in the external field between the first and last interactions with the target. Each of these interactions can occur on the quark or on the antiquark, hence the four terms. One contribution is depicted in~\fign{fig:operator}. Using $A^+(\x_2)-A^+(\y_2)=\int_{\y_2}^{\x_2}\rmd z^i F^{i+}(z)$ allows to combine and simplify all contributions. \\
It is worth noting that the wave functions are unmodified by the inclusion of presumably $1/\sqrt{s}$-suppressed terms into the shock wave picture. The notion of the perceived extent of the target is very natural here: it can be understood as the time difference between the first and the last interaction. In the shock wave approximation, this difference would be assumed to be close to 0 and an expansion around that point would be performed~\cite{next-to-eikonal,Chirilli:2018kkw}.


Let us turn now to our third approximation, the modified shock wave. When considering quantum diffusion in the external field, the propagator between interactions at points $x_i,x_j$ is a Gaussian with the exponent $ -i \ell_1^-\x_{ij}^2/(2 x_{ij}^-)$\footnote{Double subscript denote differences, e.g. $ \x_{ij} \equiv \x_i-\x_j$}. Parametrically, $x_{ij}^- \sim 1/P^+$ and $\ell^- \sim q^-$, which means $\x_{ij}^2 \sim x_{\rm Bj}/Q^2$.  In other words, the main contribution for the Brownian motion through the external field conserves the transverse position, up to corrections which are suppressed in both the Regge and the Bjorken limit. We can thus perform part of the classical expansion from~\cite{next-to-eikonal} in a way which is consistent for both small and large $x_{\rm Bj}$. We readily find
\beq
&&(\x_2 | \cG_{zq^-}(x_2^-,x_1^-) |\x_1) = \\
 &&\qquad\cG^{(0)}_{zq^-}(x_{21}^-,\x_{21}) [x_2^-,x_1^-]_\b +\cO( |\x_{21}|/{|\b |} )\nn
&&(\x_1+\r| \cG_{-\bar zq^-}(x_1^-,x_2^-) |\x_2+\r^\prime) = \\
&&\qquad \cG^{(0)}_{-\bar zq^-}(x_{12}^-,\v) [x_1^-,x_2^-]_{\b+{\boldsymbol R}}+\cO( |\v |/|\b+{\boldsymbol R}|   ) \nonumber.
\eeq
Here, $\b = (\x_1+\x_2)/2$, $\v = \x_{12} +\r-\r^\prime $ and ${\boldsymbol R} = (\r+\r^\prime)/2$.
For the same parametric reasons, we can expand the arguments of the gluon fields around $\x_1 \simeq \x_2 \simeq \b$, $\r \simeq \r^\prime \simeq {\boldsymbol{R}}$. Going to momentum space and integrating out $\b$ and $\boldsymbol{R}$ yields our final result for the cross section, for photon helicity $\lambda = L,+1,-1$\footnote{For simplicity, we do not consider spin asymmetries in this expressions. In more complicated cases, $x\leftrightarrow -x$ and $ \k \leftrightarrow -\k$ non-symmetries would lead to more cumbersome, less symmetrical, distributions than $G^{ij}$.}:
 \bwt
 \begin{equation}  \label{eq:cross-section}
 \sigma_{\lambda}(x_{\rm Bj}, Q^2)= 4 \alpha_{\rm em}\alpha_s \sum_f q_f^2 \! \int_0^1 \! \frac{\rmd x}{2\pi}\int_0^1 \! \frac{\rmd z}{2\pi} \int \! \rmd^2 \k \, \rmd^2 \bell \, \del^i \phi_\lambda \! \left(\bell+\frac{\k}{2}\right)\del^j \phi_{\lambda}^\ast \! \left(\bell-\frac{\k}{2}\right) \! \delta \! \left( x -x_{Bj}-\frac{\bell^2}{2z\bar z q^-P^+}\right) \!  x  \, G^{ij}(x,\k),
 \end{equation}
 \ewt
 where the unintegrated gluon distribution is defined in \eqn{eq:dist-def} and the wave functions $\phi_\lambda$ are the Fourier transforms of the $\varphi_\lambda$ functions from \eqn{eq:phiL}, \eqn{eq:phiT}. In terms of the original variables $\k=\bell_2-\bell_1$ and $\bell=(\bell_2+\bell_1)/2$. This equation is exact up to corrections of relative order $p_\perp/\sqrt{s}$, where $p_\perp$ is an intrinsic transverse momentum in the proton. Such corrections are suppressed in the Bjorken regime as well as in the Regge limit. The explicit $x$ dependence in 
Eqs.~(\ref{eq:cross-section}) and (\ref{eq:cross-section-position}) results in a non-locality in transverse dipole sizes which is not compatible with the dipole model as previously noted in \cite{Bialas:2000xs}. Only when $x$ is neglected do we get $\delta(\r-\r')$. 
\eqn{eq:cross-section} together with  \eqn{eq:dist-def} is our main result. 

Finally, it is worth noting that when $|\k| \ll |\bell|$, \eqn{eq:cross-section} reproduces the DGLAP logarithm associated with the creation of the quark (resp. antiquark) from the splitting of a collinear gluon in the target. This limit is achieved for $z \simeq 0$ (resp. $z \simeq 1$). Integrating over $z= y/(1-y)\bell^2/Q^2$ using the delta function in \eqn{eq:cross-section} with $y=x_{Bj}/x$ and neglecting $\k$ in the wave functions one recovers the gluon PDF upon integration over $\k$, multiplied by the Altarelli-Parisi splitting function $P_{qg}(y) \sim y^2+(1-y)^2$ and $\int^{Q^2}_{\mu^2} \rmd \bell^2/\bell^2 = \log Q^2/\mu^2 $. 

\section{Summary and outlook}
By investigating the small-$x$-inspired semi-classical description of an observable beyond the ``naive'' high energy limit, aka the shock wave approximation, we found an unintegrated gluon distribution with explicit dependence on the longitudinal fraction which spans both the collinear and the small $x$ limits. This distribution, remarkably, does not involve infinite-length Wilson lines, hence its evaluation on the lattice is greatly simplified when compared to the usual small-$x$ distributions. Such a lattice study would directly confirm or infirm the existence and energy dependence~\cite{Mueller:2002zm, Munier:2003vc} of the saturation scale from first principles.

Phenomenology for semi-classical small-$x$ physics at NLL accuracy has revealed a consistency issue with the standard   scheme~\cite{negativeXS}. Colossal efforts were made in order to address it, mostly by modifying the evolution equation without changing the evolved quantity~\cite{collinear-logs-Beuf,collinear-logs-Edmond}. The origin of this inconsistency is easy to identify in \eqn{eq:cross-section}. Indeed, it is widely believed that the smallness of $x_{\rm Bj }$ leads to the smallness of the longitudinal fraction $x$. This fraction can however be enhanced by loop integrals even at small values of $x_{\rm Bj}$: the $z\rightarrow 0,1$ limits in \eqn{eq:cross-section} yield the full DGLAP logarithms when $\bell \gg \k$. By requiring $x\simeq x_{\rm Bj} \simeq 0$, the shock wave approximation leads to an inconsistent treatment of such collinear logarithms. Studying the quantum evolution of the new distribution derived in this article should provide an alternative to the BK equation and solve this conundrum in a natural way. \\
In a similar fashion, the limit $x \simeq 0$ tends to suppress target spin effects~\cite{Hatta:2016aoc,Kovchegov:2017lsr}. We can actually conclude from the present analysis that spin effects can occur even in the small $x_{\rm Bj}$ limit because of the collinear corner of the phase space where $x\gg x_{\rm Bj}$. \\

\begin{acknowledgements}
\section{Acknowledgements}
This work was supported by the U.S. Department of Energy, Office of Science, Office of Nuclear Physics, under contract No. DE- SC0012704, and by Laboratory Directed Research and Development (LDRD) funds from Brookhaven Science Associates. Y. M.-T. acknowledges support from the RHIC Physics Fellow Program of the RIKEN BNL Research Center. Figures were drawn using Jaxodraw.
\end{acknowledgements}

\bibliographystyle{apsrev4-1}

\begin{thebibliography}{4}%
\makeatletter
\providecommand \@ifxundefined [1]{%
 \@ifx{#1\undefined}
}%
\providecommand \@ifnum [1]{%
 \ifnum #1\expandafter \@firstoftwo
 \else \expandafter \@secondoftwo
 \fi
}%
\providecommand \@ifx [1]{%
 \ifx #1\expandafter \@firstoftwo
 \else \expandafter \@secondoftwo
 \fi
}%
\providecommand \natexlab [1]{#1}%
\providecommand \enquote  [1]{``#1''}%
\providecommand \bibnamefont  [1]{#1}%
\providecommand \bibfnamefont [1]{#1}%
\providecommand \citenamefont [1]{#1}%
\providecommand \href@noop [0]{\@secondoftwo}%
\providecommand \href [0]{\begingroup \@sanitize@url \@href}%
\providecommand \@href[1]{\@@startlink{#1}\@@href}%
\providecommand \@@href[1]{\endgroup#1\@@endlink}%
\providecommand \@sanitize@url [0]{\catcode `\\12\catcode `\$12\catcode
  `\&12\catcode `\#12\catcode `\^12\catcode `\_12\catcode `\%12\relax}%
\providecommand \@@startlink[1]{}%
\providecommand \@@endlink[0]{}%
\providecommand \url  [0]{\begingroup\@sanitize@url \@url }%
\providecommand \@url [1]{\endgroup\@href {#1}{\urlprefix }}%
\providecommand \urlprefix  [0]{URL }%
\providecommand \Eprint [0]{\href }%
\providecommand \doibase [0]{http://dx.doi.org/}%
\providecommand \selectlanguage [0]{\@gobble}%
\providecommand \bibinfo  [0]{\@secondoftwo}%
\providecommand \bibfield  [0]{\@secondoftwo}%
\providecommand \translation [1]{[#1]}%
\providecommand \BibitemOpen [0]{}%
\providecommand \bibitemStop [0]{}%
\providecommand \bibitemNoStop [0]{.\EOS\space}%
\providecommand \EOS [0]{\spacefactor3000\relax}%
\providecommand \BibitemShut  [1]{\csname bibitem#1\endcsname}%
\let\auto@bib@innerbib\@empty
\bibitem [{Note1()}]{Note1}%
  \BibitemOpen
  \bibinfo {note} {It is customary in collider physics to use light cone
  variables defined as $k^+=(k_0+k_3)/\protect \sqrt {2}$ and
  $k^-=(k_0-k_3)/\protect \sqrt {2}$ in addition to the transverse components
  ${\protect \boldsymbol k}\equiv (k^1,k^2)$. Hence, in the frame where the
  proton is moving in the $+z$ direction its dominant momentum component is
  $P^+$.}\BibitemShut {Stop}%
\bibitem [{Note2()}]{Note2}%
  \BibitemOpen
  \bibinfo {note} {$F^{i+}(\xi ^-,{\protect \boldsymbol r})\equiv F^{i+}(\xi
  ^-,\xi ^+=0,{\protect \boldsymbol r})$}\BibitemShut {NoStop}%
\bibitem [{Note3()}]{Note3}%
  \BibitemOpen
  \bibinfo {note} {Double subscript denote differences, e.g. $ {\protect
  \boldsymbol x}_{ij} \equiv {\protect \boldsymbol x}_i-{\protect \boldsymbol
  x}_j$}\BibitemShut {NoStop}%
\bibitem [{Note4()}]{Note4}%
  \BibitemOpen
  \bibinfo {note} {For simplicity, we do not consider spin asymmetries in this
  expressions. In more complicated cases, $x\leftrightarrow -x$ and $ {\protect
  \boldsymbol k}\leftrightarrow -{\protect \boldsymbol k}$ non-symmetries would
  lead to more cumbersome, less symmetrical, distributions than
  $G^{ij}$.}\BibitemShut {Stop}%
\end{thebibliography}%


\begin{thebibliography}{9}

\bibitem{glr}
L.~Gribov, E.~Levin and M.~Ryskin,
Phys. Rept. \textbf{100}, 1-150 (1983)
doi:10.1016/0370-1573(83)90022-4.

\bibitem{mq}
A.~H.~Mueller and J.~w.~Qiu,
Nucl. Phys. B \textbf{268}, 427-452 (1986)
doi:10.1016/0550-3213(86)90164-1.

\bibitem{dglap}
V.~Gribov and L.~Lipatov,
Sov. J. Nucl. Phys. \textbf{15}, 438-450 (1972)
IPTI-381-71 ; 
Y.~L.~Dokshitzer,
Sov. Phys. JETP \textbf{46}, 641-653 (1977) ;
G.~Altarelli and G.~Parisi,
Nucl. Phys. B \textbf{126}, 298-318 (1977)
doi:10.1016/0550-3213(77)90384-4.

\bibitem{negativeXS}
T.~Lappi and H.~M\"{a}ntysaari,
Phys. Rev. D \textbf{91}, no.7, 074016 (2015)
doi:10.1103/PhysRevD.91.074016
[arXiv:1502.02400 [hep-ph]] ; 
E.~Avsar, A.~Stasto, D.~Triantafyllopoulos and D.~Zaslavsky,
JHEP \textbf{10}, 138 (2011)
doi:10.1007/JHEP10(2011)138
[arXiv:1107.1252 [hep-ph]].

\bibitem{nlobfkl}
V.~S.~Fadin and L.~Lipatov,
Phys. Lett. B \textbf{429}, 127-134 (1998)
doi:10.1016/S0370-2693(98)00473-0
[arXiv:hep-ph/9802290 [hep-ph]] ; M.~Ciafaloni and G.~Camici,
Phys. Lett. B \textbf{430}, 349-354 (1998)
doi:10.1016/S0370-2693(98)00551-6
[arXiv:hep-ph/9803389 [hep-ph]].

\bibitem{nlobk}
I.~Balitsky and G.~A.~Chirilli,
Phys. Rev. D \textbf{77}, 014019 (2008)
doi:10.1103/PhysRevD.77.014019
[arXiv:0710.4330 [hep-ph]].

\bibitem{nlojimwlk}
A.~Kovner, M.~Lublinsky and Y.~Mulian,
Phys. Rev. D \textbf{89}, no.6, 061704 (2014)
doi:10.1103/PhysRevD.89.061704
[arXiv:1310.0378 [hep-ph]] ; A.~Kovner, M.~Lublinsky and Y.~Mulian,
JHEP \textbf{08}, 114 (2014)
doi:10.1007/JHEP08(2014)114
[arXiv:1405.0418 [hep-ph]].

\bibitem{bfkl}
I.~Balitsky and L.~Lipatov, ``{The Pomeranchuk Singularity in Quantum
  Chromodynamics},''
{\em Sov.J.Nucl.Phys.} {\bf 28} (1978)  822--829 ; E.~A.~Kuraev, L.~N.~Lipatov and V.~S.~Fadin,
Sov. Phys. JETP \textbf{44}, 443-450 (1976) ; 
E.~Kuraev, L.~Lipatov, and V.~S. Fadin, ``{The Pomeranchuk Singularity in
  Nonabelian Gauge Theories},''
{\em Sov.Phys.JETP} {\bf 45} (1977)  199--204. 

\bibitem{bkian}
I.~Balitsky,
Nucl. Phys. B \textbf{463} (1996), 99-160
doi:10.1016/0550-3213(95)00638-9
[arXiv:hep-ph/9509348 [hep-ph]].

\bibitem{bkyuri}
Y.~V.~Kovchegov,
Phys. Rev. D \textbf{60} (1999), 034008
doi:10.1103/PhysRevD.60.034008
[arXiv:hep-ph/9901281 [hep-ph]].

\bibitem{jimwlk}
J.~Jalilian-Marian, A.~Kovner, A.~Leonidov and H.~Weigert,
Phys. Rev. D \textbf{59} (1998), 014014
doi:10.1103/PhysRevD.59.014014
[arXiv:hep-ph/9706377 [hep-ph]] ; E.~Iancu, A.~Leonidov and L.~D.~McLerran,
Nucl. Phys. A \textbf{692} (2001), 583-645
doi:10.1016/S0375-9474(01)00642-X
[arXiv:hep-ph/0011241 [hep-ph]].

\bibitem{collinear-logs-Beuf}
G.~Beuf,
Phys. Rev. D \textbf{89}, no.7, 074039 (2014)
doi:10.1103/PhysRevD.89.074039
[arXiv:1401.0313 [hep-ph]].

\bibitem{collinear-logs-Edmond}
E.~Iancu, J.~Madrigal, A.~Mueller, G.~Soyez and D.~Triantafyllopoulos,
Phys. Lett. B \textbf{744}, 293-302 (2015)
doi:10.1016/j.physletb.2015.03.068
[arXiv:1502.05642 [hep-ph]] ; 
E.~Iancu, J.~Madrigal, A.~Mueller, G.~Soyez and D.~Triantafyllopoulos,
Phys. Lett. B \textbf{750}, 643-652 (2015)
doi:10.1016/j.physletb.2015.09.071
[arXiv:1507.03651 [hep-ph]] ; B.~Duclou\'{e}, E.~Iancu, A.~Mueller, G.~Soyez and D.~Triantafyllopoulos,
JHEP \textbf{04}, 081 (2019)
doi:10.1007/JHEP04(2019)081
[arXiv:1902.06637 [hep-ph]] ; B.~Duclou\'{e}, E.~Iancu, G.~Soyez and D.~Triantafyllopoulos,
Phys. Lett. B \textbf{803}, 135305 (2020)
doi:10.1016/j.physletb.2020.135305
[arXiv:1912.09196 [hep-ph]].

\bibitem{mv}
L.~D.~McLerran and R.~Venugopalan,
Phys. Rev. D \textbf{49}, 2233-2241 (1994)
doi:10.1103/PhysRevD.49.2233
[arXiv:hep-ph/9309289 [hep-ph]] ; L.~D.~McLerran and R.~Venugopalan,
Phys. Rev. D \textbf{49}, 3352-3355 (1994)
doi:10.1103/PhysRevD.49.3352
[arXiv:hep-ph/9311205 [hep-ph]] ; L.~D.~McLerran and R.~Venugopalan,
Phys. Rev. D \textbf{50}, 2225-2233 (1994)
doi:10.1103/PhysRevD.50.2225
[arXiv:hep-ph/9402335 [hep-ph]].

\bibitem{updf}
K.~Kutak and A.~Stasto,
Eur. Phys. J. C \textbf{41}, 343-351 (2005)
doi:10.1140/epjc/s2005-02223-0
[arXiv:hep-ph/0408117 [hep-ph]].

\bibitem{Munier:2003vc}
S.~Munier and R.~B.~Peschanski,
Phys. Rev. Lett. \textbf{91}, 232001 (2003)
doi:10.1103/PhysRevLett.91.232001
[arXiv:hep-ph/0309177 [hep-ph]].

\bibitem{Kovchegov:1999ua}
Y.~V.~Kovchegov,
Phys. Rev. D \textbf{61} (2000), 074018
doi:10.1103/PhysRevD.61.074018
[arXiv:hep-ph/9905214 [hep-ph]].

\bibitem{Balitsky:2015qba}
I.~Balitsky and A.~Tarasov,
JHEP \textbf{10}, 017 (2015)
doi:10.1007/JHEP10(2015)017
[arXiv:1505.02151 [hep-ph]].

\bibitem{GMT}
F.~Gelis and Y.~Mehtar-Tani,
Phys. Rev. D \textbf{73} (2006), 034019
doi:10.1103/PhysRevD.73.034019
[arXiv:hep-ph/0512079 [hep-ph]].

\bibitem{next-to-eikonal}
T.~Altinoluk, N.~Armesto, G.~Beuf, M.~Martínez and C.~A.~Salgado,
JHEP \textbf{07}, 068 (2014)
doi:10.1007/JHEP07(2014)068
[arXiv:1404.2219 [hep-ph]] ;  T.~Altinoluk, N.~Armesto, G.~Beuf and A.~Moscoso,
JHEP \textbf{01}, 114 (2016)
doi:10.1007/JHEP01(2016)114
[arXiv:1505.01400 [hep-ph]].

\bibitem{Chirilli:2018kkw}
G.~A.~Chirilli,
JHEP \textbf{01}, 118 (2019)
doi:10.1007/JHEP01(2019)118
[arXiv:1807.11435 [hep-ph]].

\bibitem{Blaizot:2015lma}
J.~P.~Blaizot and Y.~Mehtar-Tani,
Int. J. Mod. Phys. E \textbf{24} (2015) no.11, 1530012
doi:10.1142/S021830131530012X
[arXiv:1503.05958 [hep-ph]].

\bibitem{Beuf:2011xd}
G.~Beuf,
Phys. Rev. D \textbf{85}, 034039 (2012)
doi:10.1103/PhysRevD.85.034039
[arXiv:1112.4501 [hep-ph]].

\bibitem{Bialas:2000xs}
A.~Bialas, H.~Navelet and R.~B.~Peschanski,
Nucl. Phys. B \textbf{593} (2001), 438-450
doi:10.1016/S0550-3213(00)00640-4
[arXiv:hep-ph/0009248 [hep-ph]].

\bibitem{Mueller:2002zm}
A.~Mueller and D.~Triantafyllopoulos,
Nucl. Phys. B \textbf{640}, 331-350 (2002)
doi:10.1016/S0550-3213(02)00581-3
[arXiv:hep-ph/0205167 [hep-ph]].

\bibitem{Hatta:2016aoc}
Y.~Hatta, Y.~Nakagawa, F.~Yuan, Y.~Zhao and B.~Xiao,
Phys. Rev. D \textbf{95}, no.11, 114032 (2017)
doi:10.1103/PhysRevD.95.114032
[arXiv:1612.02445 [hep-ph]].

\bibitem{Kovchegov:2017lsr}
Y.~V.~Kovchegov, D.~Pitonyak and M.~D.~Sievert,
JHEP \textbf{10}, 198 (2017)
doi:10.1007/JHEP10(2017)198
[arXiv:1706.04236 [nucl-th]].



\end{thebibliography}

\end{document}